\documentclass[12pt]{article}
\usepackage{rotating}
\usepackage{booktabs, array, dcolumn}
\usepackage{multirow}
\usepackage[hyphens]{url}
\usepackage[hidelinks]{hyperref}
\hypersetup{breaklinks=true}

\title{Blockchain-based Smart Contracts - Applications and Challenges} 

\author{
	Yining Hu \\
	University of New South Wales and Data61-CSIRO
	\and
	Madhusanka Liyanage\\
	University College Dublin
	\and
	Ahsan Manzoor\\
	Rovio Entertainment
	\and
	Kanchana Thilakarathna\\
	University of Sydney
	\and
	Guillaume Jourjon\\
	Data61-CSIRO
	\and
	Aruna Seneviratne\\
	University of New South Wales
}

\begin{document}
\maketitle

\begin{abstract}
A blockchain-based smart contract or a "smart contract" for short, is a computer program intended to digitally facilitate the negotiation or contractual terms directly between users when certain conditions are met. With the advance in blockchain technology, smart contracts are being used to serve a wide range of purposes ranging from self-managed identities on public blockchains to automating business collaboration on permissioned blockchains. In this paper, we present a comprehensive survey of smart contracts with a focus on existing applications and challenges they face.
\end{abstract}

\section{Introduction}\label{sec:intro}
\subsection{What Are Smart Contracts?}
\begin{figure}[!ht]
	\centering
	\includegraphics[scale=.35]{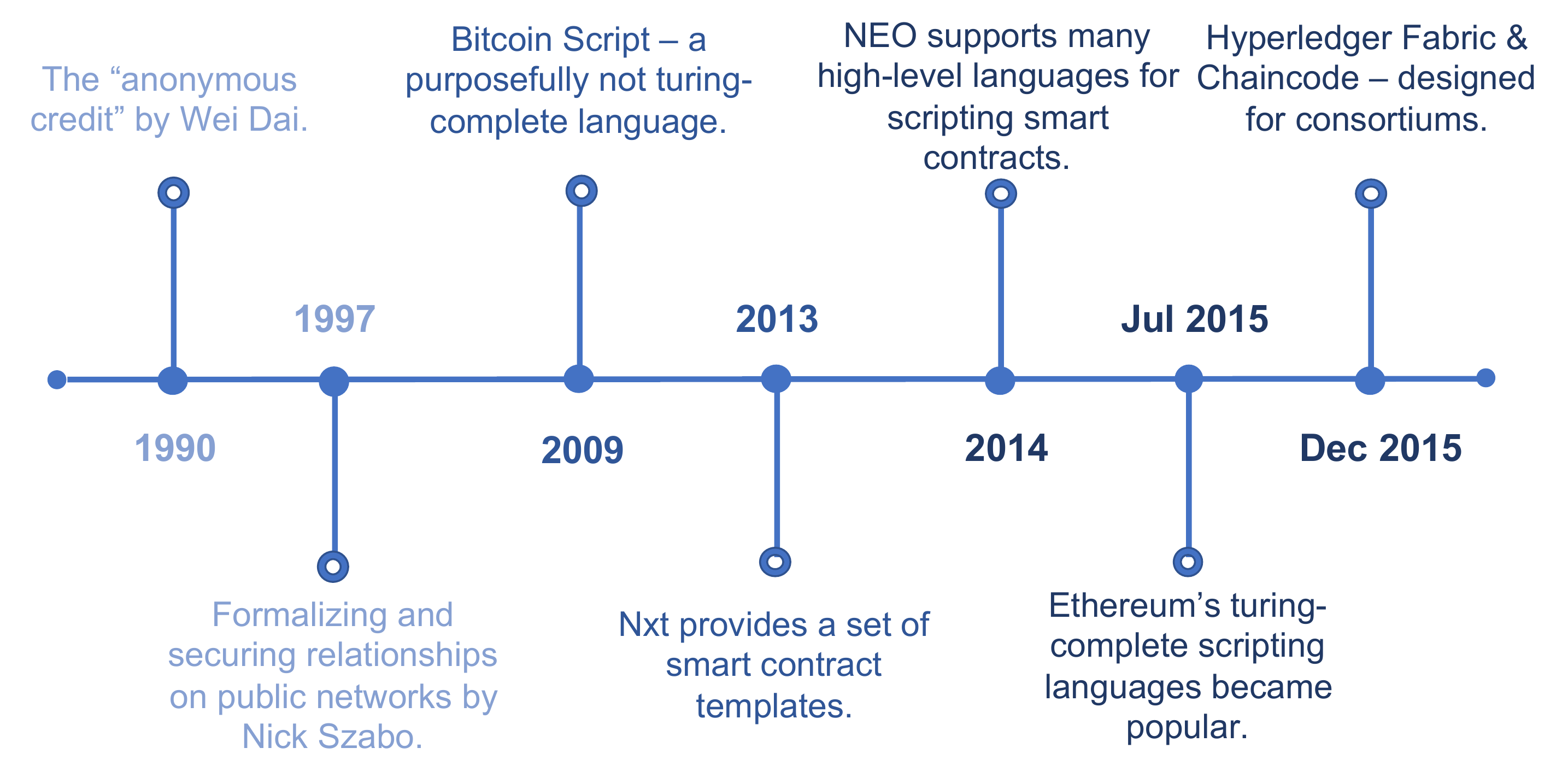}
	\caption{Evolution of smart contracts.} 
	\label{fig:sc-evolution}
\end{figure}

The history of smart contracts can be traced back to the 1990s when Wei Dai, a computer engineer created a post on anonymous credits, which described an anonymous loan scheme with redeemable bonds and lump-sum taxes to be collected at maturity~\cite{weidai_post}. Szabo et al.~\cite{szabo1997formalizing} later discussed the potential form of smart contracts and proposed to use cryptographic mechanisms to enhance security.
Nowadays, with the development of blockchain technology, smart contracts are being constructed as computer programs running on blockchain nodes and can be issued among untrusted, anonymous parties without the involvement of any third party.
The first successful implementation of a blockchain-based smart contract was Bitcoin Script~\cite{btc_script}, a purposely not-turing-complete language with a set of simple, pre-defined commands. As simple forms of smart contract, standard types of Bitcoin transactions, such as pay-to-public-key-hash (P2PKH) and pay-to-script-hash (P2SH), are all defined with Bitcoin Script~\cite{antonopoulos2014mastering}.
In addition, there also exist platforms that enable more complex contractual functionalities and flexibilities, e.g., Ethereum~\cite{wood2014ethereum}, which adopts a turing-complete language for smart contracts. 
Newer blockchain platforms such as Neo~\cite{neo} and Hyperledger Fabric~\cite{hyperledger} allow smart contracts to be written in various high-level languages. Figure~\ref{fig:sc-evolution} illustrates the evolution of smart contracts.

\subsection{Why Do We Need Smart Contracts?}
Smart contracts inherit properties of underlying blockchains which include an immutable record of data, and the ability to mitigate single points of failure. 
Smart contracts can also interact with each other via calls.
Unlike traditional paper contracts that rely on middlemen and third-party intermediaries for execution, smart contracts automate contractual procedures, minimize interactions between parties, and reduce administration cost. 

Due to the ease of deployment, smart contracts on public blockchains or "public smart contracts (cf. Section~\ref{sec:mechanism}) have attracted a wide variety of commercial applications. While smart contracts on permissioned blockchains or "permissioned smart contracts" are more often used in collaborative business processes (cf. Section~\ref{sec:mechanism}) since they have the potential to \emph{prevent unwanted updates}, \emph{improve efficiency} and \emph{save costs}.

\begin{table}[!ht]
	\tiny
	\centering
	\begin{tabular}{|p{2cm}|p{4cm}|p{4cm}|}\specialrule{.12em}{1em}{0em}
		& \textbf{Public Smart Contracts} & \textbf{Permissioned Smart Contracts} \\ \hline 
		
		\centering{\textbf{Common}} & \multicolumn{2}{c|}{Immutable record} \\
		& \multicolumn{2}{c|}{Proper encyption on data and pseudonymity} \\
		& \multicolumn{2}{c|}{Interoperability among different platforms} \\
		& \multicolumn{2}{c|}{Traceable modifications} \\ \hline
		
		\centering{\textbf{Unique}} & Easy to deploy \newline Accessible for the public & Faster settlement \newline Lower operational cost \newline Permissioned access \\ \hline
		
		\specialrule{.12em}{0em}{0em}
	\end{tabular}
	\caption{Characteristics of public and permissioned smart contracts.}
	\label{tab:characteristics}
\end{table}

Despite the hype of blockchain and smart contracts, the technology is still in its infancy. This paper explores the differences between public and permissioned smart contracts, provides examples for existing smart contract applications, discusses existing research and highlights remaining challenges to overcome for a fuller adoption of the technology. Different than existing research that classifies smart contracts based on their application areas~\cite{casino2018systematic} or only discusses the technical aspect of smart contracts~\cite{wang2019blockchain}, we classify smart contracts into public and permissioned and look into the legal aspect and usability of smart contracts.

\section{Smart Contract Mechanisms}
\label{sec:mechanism}
\subsection{Overview}
The operation of smart contracts can hardly be decoupled from the underlying blockchain. 
State of a blockchain is updated when a valid transaction is recorded on chain~\cite{bonneau2015sok}, and smart contracts can be used to automatically trigger transactions under certain conditions.
We categorize smart contracts to public smart contracts and permissioned smart contracts according to the blockchain platforms they operate on. As the expectation and requirements for smart contracts are often different for the two categories, we below discuss them separately. We consider all smart contracts on permissioned, consortium or private blockchains as permissioned smart contracts.

\subsection{Public Smart Contracts}
\begin{figure}[!ht]
	\centering
	\includegraphics[scale=.6]{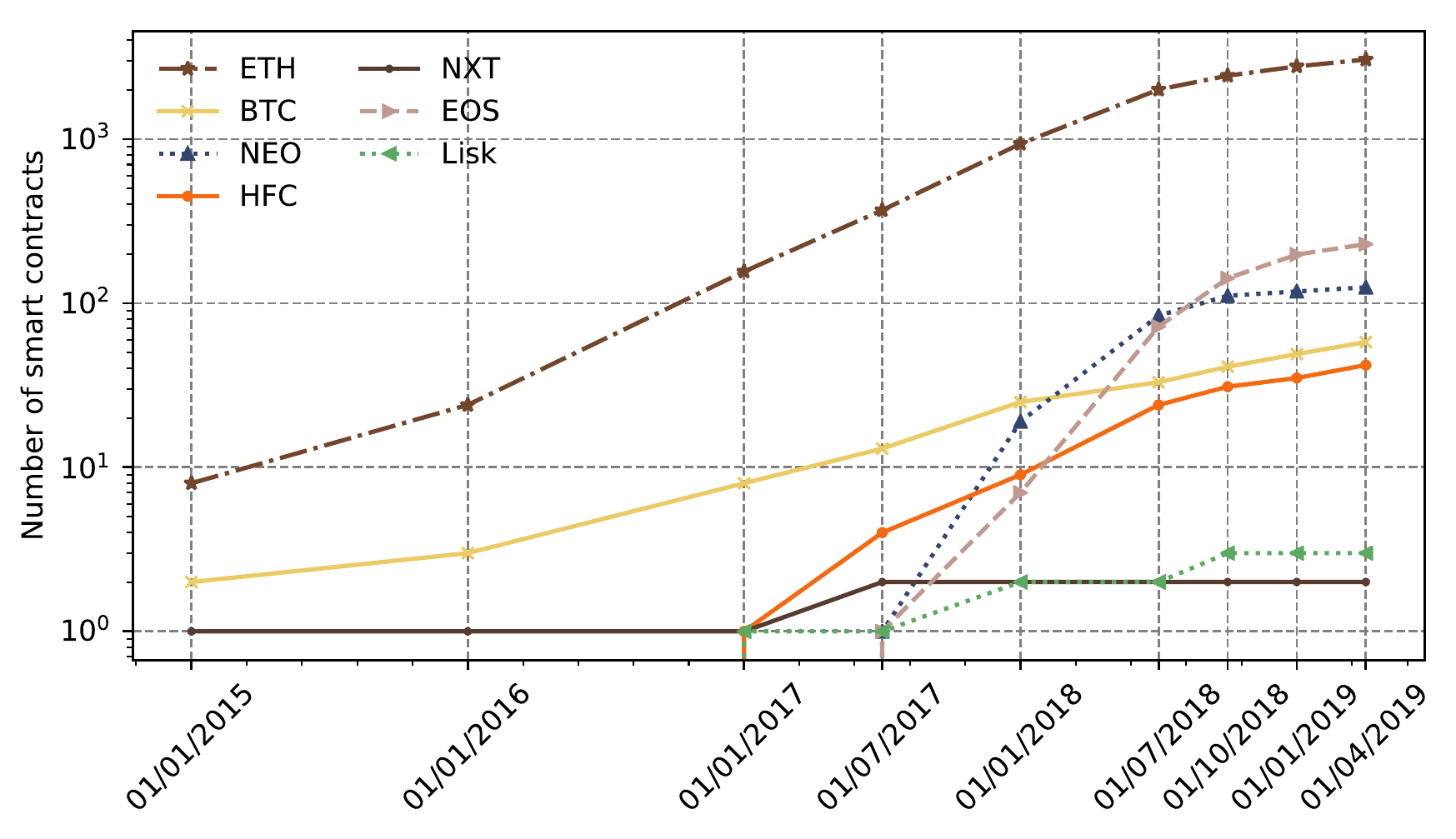}
	\caption{Number of smart contracts on popular blockchains.} 
	\label{fig:num-of-sc}
\end{figure}

Public blockchains set no requirement for peers to participate, hence all peers have the right to deploy smart contracts. In order to prevent spamming, when instantiating or invoking smart contracts on a public blockchain, one is often required to pay a certain amount of fee. Limited by it's functionality, the scripting language used in Bitcoin--Scripts~\cite{btc_script}--is hardly used in constructing complex contractual terms. While the general-purpose Solidity language~\cite{solidity} in Ethereum can be used for a much wider variety of applications.
According to Etherscan~\cite{etherscan}, among the one million Ethereum accounts that altogether hold 105.6 million Ethers,\footnote{This equals 19.1 billion USD at the time of writing.} half of them are contract accounts with a total balance of 12 million Ether.
Competitors such as Neo~\cite{neo} and EOS~\cite{eosio}, are also independent blockchains facilitating peer consensus and smart contracts. 
To show the popularity of different platforms, we obtained the number of publicly available smart contract projects deployed on Github~\cite{github} from the beginning of 2015 till early 2019. As illustrated in Figure~\ref{fig:num-of-sc}, Ethereum is the most popular platform among the 7 blockchain instances we surveyed.

To give readers an intuitive idea of how smart contracts work on public blockchains, we below explain the mechanism of Ethereum contracts. Ethereum uses proof-of-work (PoW) mining protocol for network consensus. Ethereum smart contracts reside in Ethereum Virtual Machines (EVMs), which isolates them from the blockchain network to prevent the code running inside from interfering with other processes. Once deployed, the smart contract obtains a unique address that is linked to a balance, similar to an externally controlled account (EOA) owned by a user. A smart contract can send transactions to an EOA or another contract.

\begin{figure}[!ht]
	\centering
	\includegraphics[scale=.5]{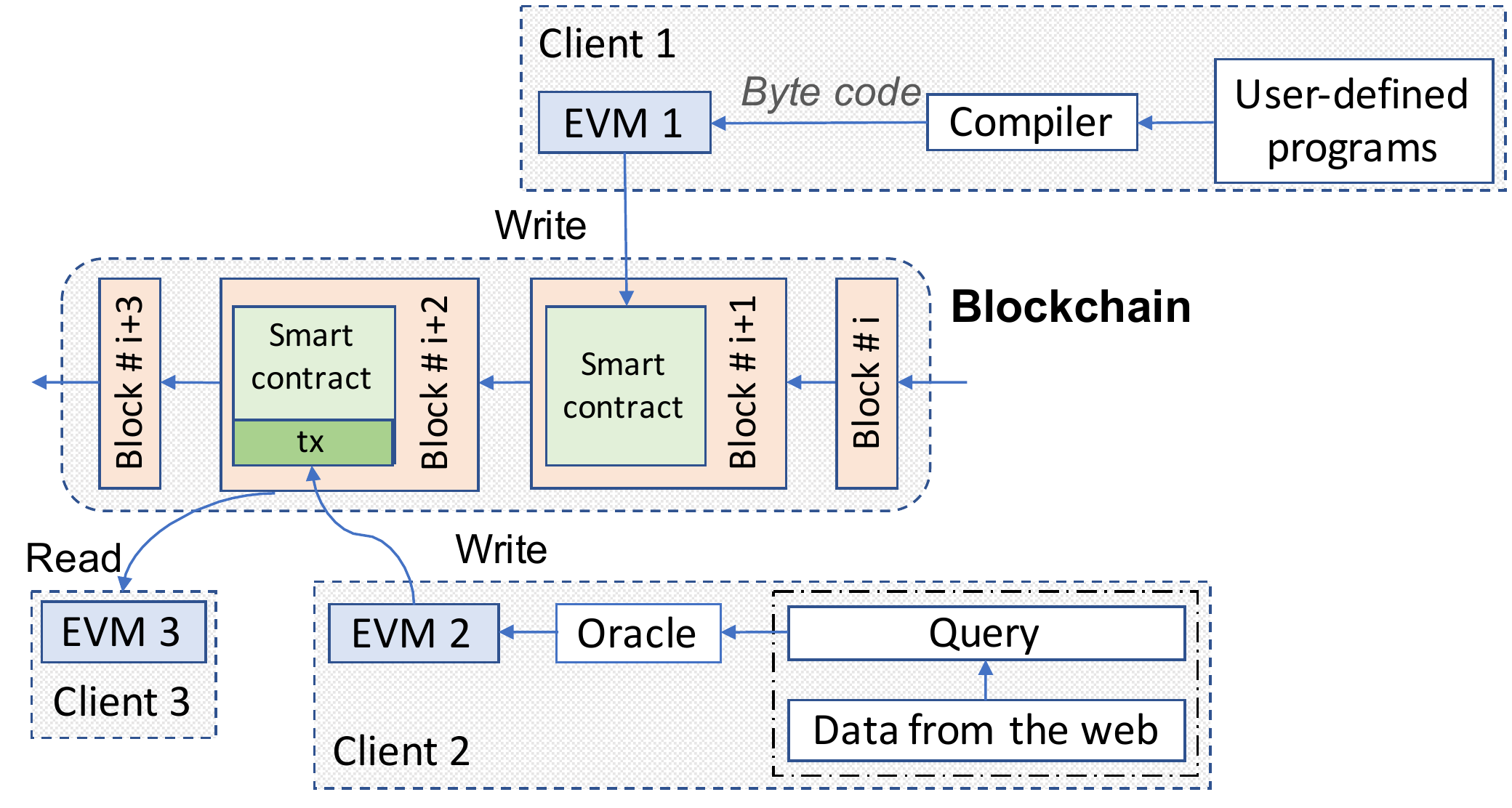}
	\caption{Mechanism of Ethereum smart contracts.} 
	\label{fig:mechanism}
\end{figure}

Figure \ref{fig:mechanism} illustrates the working of Ethereum smart contracts, where the mining process is omitted for simplification. In Step 1, Client 1 creates a smart contract for voting in a high-level language, e.g. Solidity~\cite{solidity}. This smart contract is compiled into machine-level byte code where each byte represents an operation, and then uploaded to the blockchain in the form of a transaction by EVM 1. A miner picks it up and confirms it in Block \#i+1. Once a voter has submitted his vote via the web interface, the EVM 2 queries the data from the web and embeds it into Transaction \emph{tx} and deploy it to the blockchain. State of the voting contract is updated in Block \#i+2 with the confirmation of transaction \emph{tx}. If Client 3, the coordinator, later wants to check the states stored in the contract, s/he has to synchronize up to at least Block \#i+2 to see the changes caused by \emph{tx}. 

\subsection{Permissioned Smart Contracts}
Permissioned smart contracts, residing on permissioned blockchains are becoming increasingly popular in business collaborations. 
Compared to the inefficient and expensive validation processes of public blockchains, permissioned blockchains are more suitable in stimulating business collaborations.

As an example, the Hyperledger project~\cite{hyperledger}, primarily driven by the Linux Foundation, aims to improve business processes and collaborations that involve multiple parties. Among the collection of projects in Hyperledger, Fabric serves a foundation. 
Compared to public PoW blockchains, Fabric reduces the cost of consensus by implementing a Practical Byzantine Fault-tolerant (PBFT) protocol \cite{castro2002practical}, and leveraging channels for parallel and secure transaction processing. Channels allow participants to form virtual groups and keep their independent ledgers that are invisible to other channels. Channels provide the flexibility for business consortium to securely share information only to relevant parties.

On a Fabric network, transaction ordering is handled by a central \emph{orderer} that collects transactions submitted by \emph{committers} and takes votes from \emph{endorsers} for permanently recording transactions in blocks. The block size can be customized in either number of transactions or time of waiting.
Chaincode is the equivalence of smart contracts in Hyperledger~\cite{hyperledger}. 
All participating peers are required to execute all transactions and smart contracts individually for synchronization. The IBM blockchain~\cite{hyperledger_ibm} is built on top of Fabric.

In addition, to further reduce the burden of blockchain peers, some suggest that complex business logics should be moved to a separate middle layer beyond the blockchain. For instance, Microsoft Azure is developing Cryptlets~\cite{bletchley}, where a central host executes smart contracts to support the separation of data and logic on permissioned blockchains. 

\section{Smart Contract Applications} \label{sec:applications}
\subsection{Public Smart Contracts}
Public blockchains enable convenient development and testing of smart contract applications or decentralized apps (D-Apps). Public smart contracts make it possible for startups to raise funds through Initial Coin Offerings (ICOs)~\cite{ico}. 
Big enterprises on the other hand, mainly want to take the advantage of permissioned smart contracts for incorporating their models and enforcing business procedures. Some of the popular use cases include: banking, Electronic Medical Record (EMR), IoT data management~\cite{christidis2016blockchains}. In addition, there are also other interesting applications such as smart waste management, real estate, and ride-sharing arcade city. We conducted a comprehensive survey of existing smart contract applications and discuss their strengths, weaknesses, as well as their potential of a wider adoption. 

\subsubsection{Health Care and Medical Records}
One major application area of smart contracts is related to healthcare and access control of medical records. Blockchain technology and smart contracts are seen by many healthcare professionals as a secure way of sharing and accessing patients' EMR. Smart contracts can feature multi-signature approvals between patients and providers to only allow authorized users or devices to access or append the record. They also enable interoperability via collaborative version control to maintain the consistency of the record. Besides benefiting patients and their care providers, smart contracts can also be used to grant researchers access to certain personal health data and enable micro-payments to be automatically transferred to patients for participation~\cite{ablockchainforhealthcare}.

However, the realization of these applications is limited by the immature infrastructure of most public blockchains and high development costs. There are also concerns about policies and users' willingness to publicize their personal information. 

\subsubsection{Identity Management}
uPort~\cite{uport} is an identity management framework that leverages public Ethereum smart contracts to recover accounts and protect user privacy in the case of a device loss. The main component--uPort identifier--is a unique 20-byte hexadecimal string representing the address of a proxy contract that lies in-between a controller contract and an application contract. uPort enables users to replace their private key (saved off-chain) while maintaining an on-chain persistent identifier. If a valid user brings a new device, s/he can seek for approval from a list of existing recovery delegates, and replace the old user address with a new one. Similarly, Sovrin~\cite{sovrin} is a digital identity management platform built on a public blockcahin.

Identity management frameworks using blockchain still need to go through a number of enhancements before adoption. In the case of uPort, the publicity of the recovery delegates of a user poses the security risk of compromising user identities. 

\subsubsection{Scaling Blockchains}
Despite the fundamental limits in the expressiveness of Bitcoin Script~\cite{btc_script}, the simplicity of this language helps prevent malicious contracts and safeguard the system. Bitcoin has been developing the Lightning Network \cite{poon2015bitcoin} using Script to facilitate transactions in off-chain payment channels. The goal is to improve the scalability of the Bitcoin blockchain by reducing on-chain verification and storage. A similar scheme in Ethereum is the Raiden Network~\cite{raiden}.

\subsection{Permissioned Smart Contracts}
Public smart contracts imposes inevitable threats to user privacy. More sensitive business use cases such as banking, supply chain, IoT are more commonly deployed as permissioned smart contracts. We below provide discussions on some of these use cases.

\subsubsection{Banking}
Smart contracts can be used to enforcing rules and policies in banking, for example, the mortgage service. According to a report made by Capgemini Consulting \cite{capgemini}, with smart contracts in mortgage, consumers could potentially save 480-960 USD per loan, while banks would be able to cut 3-11 billion USD of annual costs in the US and Europe. Banks can also use smart contracts to streamline clearing and settlement processes. It has been reported that more than 40 global banks have participated in a consortium to test smart contracts for clearing and settlement activities~\cite{blockchainapac}. In addition, the know your customer (KYC) and anti money laundering (AML) policies can also be embedded easily with the smart contract logic. Built on top of Hyperledger Fabric, Stellar Blockchain~\cite{stellar} facilitates automatic currency exchange in International transactions.

However, the interoperability with legacy systems and the scalability of blockchains remain to be obstacles in realising such systems. Also, it is crucial that the smart contract implementation is secure against attacks that are aimed at stealing of assets or tampering of the contract code \cite{atzei2017survey}.

\subsubsection{Provenance \& Supply Chain}
Blockchain can be used to enable some of the key properties in supply chains and logistics including transparency, optimization, security and visibility of various operations in the transportation of goods \cite{sadouskaya2017adoption}. A supply chain with continuous, real-time access to reliable, shared data is more efficient than traditional supply chains. Provenance of the product via the blockchain also raises the bar on quality in production by reducing the risk of wastage and spoilage. Example use case include~\cite{korpela2017digital, abeyratne2016blockchain, tian2017supply}. 

Despite the advantages of using blockchains in supply chains, the integration of blockchains with existing platforms and business procedures is still in its early stage. The use of smart contracts for negotiating and finalizing transactions may require major changes in the supply chain workflow. Moreover, resistance from banks, exchange networks and trusted intermediaries may also delay the blockchain adoption. 

\subsubsection{Voting} 
Voting is another application that can benefit from permissioned smart contracts. A Danish political party has implemented a smart contract to ensure the fairness and transparency for internal election~\cite{danish_political}. Mccorry et al. \cite{mccorry2017smart} proposed a boardroom voting scheme that is different from existing proposals of e-voting. Mccorry's system works under the assumption of a small group of voters with known identities and provides maximum voter privacy and verifiability. Mccorry et al. have also tested the system's feasibility on a Ethereum private network and estimated the cost of 0.73 USD per voter for running it. The statistics have shown that public blockchains are more feasible for small polls whereas permissioned blockchains will be required to run national scale elections.

\subsubsection{IoT} 
A promising but controversial application scenario is the use of blockchain and smart contracts for IoT data management. Intuitively, as both systems are decentralized in nature, blockchain could be used to enhance trust in IoT systems that constantly share and exchange a large amount of data. However, the other properties of blockchain and IoT do not seem to fit naturally together. Firstly, IoT data is often sensitive, and should not be shared with everyone else. Secondly, blockchains are resource-consuming. Even with lighter consensus mechanisms, having all IoT devices to execute all programs is redundant considering their limited processing capability.

As a major player in the field, IBM is integrating the Watson IoT Platform with the IBM Blockchain built on top of Hyperledger Composer~\cite{watson_iot}. The goal is to build a trusted, low-cost and efficient business network while maintaining an indelible record to satisfy industrial and governmental requirements. Similarly, Chain of Things~\cite{chainofthings} is also trying to merge blockchain with IoT to achieve security, reliabiltiy and interoperability. 

\subsubsection{Insurance} 
In the insurance industry, smart contracts can perform error checking, routing, approve workflows, and calculate payouts based on the type of claim and the underlying policy. For example, the processing of travel insurance claims can be automatically verified against flight delays or cancellations. Smart contracts can help remove the human factor involved in the process, therefore decreasing the overall administrative cost for the insurers and increasing the transparency for the consumers \cite{capgemini}.

Nonetheless, technological limitations and legal regulations are major challenges to be addressed before shifting to smart contracts for insurance policies. Another drawback is the inflexibility of smart contracts. Traditional contracts can be amended or terminated upon agreement between both parties, but smart contracts as computer programs have no such mechanism. Moreover, more authorities are needed to recognize the legality of financial smart contacts. \\

Overall, smart contracts facilitate development of decentralized applications and have great potential to reshape business procedures. Table~\ref{tab:applications} provides descriptions for more smart contract use cases and example applications.

\begin{table}[!ht]
	\tiny
	\centering
	\begin{tabular}{|p{1.5cm}|p{2cm}|p{8cm}|}\specialrule{.12em}{1em}{0em}
		\multicolumn{2}{|c|}{\textbf{Use case}} & \textbf{The role of smart contracts} \\ \hline 
		\multirow{8}{*}{\centering{\textbf{Financial}}} & \centering{Banking} & Any possible asset, such as a fiat currency, a house or a bond, can be represented in the form of smart-contract-based tokens and consequently traded on a blockchain~\cite{peters2016understanding,eyal2017blockchain}. \\ \cline{2-3} 
		& \centering{Mortgages} & Smart contracts can provide automation, shared access to electronic versions of verified physical legal documents as well as access to external sources of information such as title deeds and land registry records~\cite{crosby2016blockchain,swan2017anticipating}. \\ \cline{2-3}
		& \centering{Trade clearing and settlements} & Smart contracts can take over the onerous administrative task of managing approvals between participants, calculating trade settlement amounts and then transferring the funds automatically once the transaction embedded within the smart contract has been verified and approved~\cite{niya2018setting,vovchenko2017competitive}. \\ \cline{2-3}
		& \centering{Know Your Customer (KYC)} & Blockchains enables a smart contract that relies on KYC information to be verified as a condition precedent automatically~\cite{schuh2015bitshares}. \\ \cline{2-3}
		& \centering{Insurance} & Smart contracts perform error checking, routing, and calculate payouts based on the type of claim and the underlying policy~\cite{giancaspro2017smart,schuh2015bitshares, cohn2017smart}. \\ \cline{2-3}
		& \centering{Bond} & Smart contracts can be used to set up and manage "smart bonds". A smart bond would mainly be in the area of permission i.e. to define detailed rules about who is allowed or not to hold the bond~\cite{savelyev2017contract}. \\ \cline{2-3}
		& \centering{Delay-tolerant micro-payments} & Smart-contract-based tokens can be used as a replacement for fiat currencies to enable payments in environments with limited or intermittent connectivity. All transactions are stored in a blockchain during the disconnection periods. Bank updates the fiat currency accounts based on the blockchain entries when connected~\cite{manzoor2018delay,hu2019delay}. \\ \cline{2-3}
		& \centering{Charity} & Smart contacts can be used to embed a geo-location signature on digital currencies in donation~\cite{savu2017quality,jayasinghe2017philanthropy}. \\ \hline
		
		\multirow{4}{*}{\textbf{Health care}} & \centering{Electronic Medical Records (EMR)} & Provides access to medical health records upon multi-signature approvals between patients and care providers~\cite{dubovitskaya2017secure,azaria2016medrec,ramani2018secure}. \\ \cline{2-3}
		& \centering{Population health data access} & Smart contracts can be used to grant access for health researchers to certain personal health information and automatically triger micro-payments to the corresponding patientss~\cite{brodersen2016blockchain,linn2016blockchain}. \\ \cline{2-3}
		& \centering{Patient matching and identification} & Smart contracts can provide a platform to share patients' information between different organizations~\cite{patel2018framework}. \\ \cline{2-3}
		& \centering{Personal health tracking} & Tracks patients' health-related actions via smart devices and automatically generates rewards based on specific milestones\cite{swan2015blockchain,engelhardt2017hitching}. \\ \hline
		
		\centering{\textbf{Identity management}} & \centering{-} & Identity management framework built with smart contract can give users direct control over their identity~\cite{dunphy2018first,jacobovitz2016blockchain,ebrahimi2017identity}. \\ \hline
		
		\centering{\textbf{Energy and resources}} & \centering{-} & Smart contracts enable the distributed agreements where users can record the excess of generated energy, such as rooftop solar energy, and sell it to other users who need it\cite{mengelkamp2018designing,goranovic2017blockchain,wang2017novel,noor2018energy}. \\ \hline
		
		\multirow{5}{*}{\centering{\textbf{Cross-industry}}} & \centering{Supply chain and trade finance} & Smart contracts can ensure proper access control for data shared among participants in the supply chain. It can be used for tracking food items from farms to packaging and shipping. Smart contracts can help identify contamination and reduce food waste in the supply chains~\cite{korpela2017digital,abeyratne2016blockchain,tian2017supply}. \\ \cline{2-3}
		& \centering{Voting} & Smart contract can validate voter criteria, log vote to the blockchain, and initiate specific actions as a result of the majority vote~\cite{frantz2016institutions,buterin2014next,mccorry2017smart}. \\ \cline{2-3}
		& \centering{Commercial Real Estate (CRE)} & The blockchain is distributed and highly availabl~e. It also retains a secure source of proof that the transaction occurred~\cite{veuger2018trust,dijkstra2017blockchain,veuger2017attention}. \\ \cline{2-3}
		& \centering{Resource-sharing} & Smart contracts enable users to register and rent devices without the involvement of a Trusted Third Party (TTP), disclosure of any personal information or prior sign-up to the service~\cite{hawlitschek2018limits,hong2017blockchain}. \\ \cline{2-3}
		& \centering{Product provenance} & Facilitates chain-of-custody process for products in the supply chain where the party in custody is able to log evidence about the product~\cite{sadouskaya2017adoption,kim2018toward}.\\ \hline
		
		\textbf{Smart city} & \centering{General} & Establish trust-free decentralized service relationships among human, technology, and organizations in a smart city~\cite{sun2016blockchain,biswas2016securing}. \\ \cline{2-3}
		&  \centering{Automotives} & A dedicated distributed ledger for automotives can track anything from the market price of a vehicle to its road safety records to its miles-per-gallon performance and so on~\cite{skulimowskiinternet,sharma2017block}.\\ \hline
		
		\multirow{2}{*}{\textbf{Technology}} & \centering{Mobile networks} & Provisions and agreements between operators, access nodes, networks, and subscribers are negotiated on-the-fly as digital smart contracts. When a device negotiates the best service, the carrier dynamically adjusts the smart contract code. Roaming agreements between a visitor and the home network can also be implemented~\cite{fan2017blockchain,backman2017blockchain,di2017smart}. \\ \cline{2-3}
		
		& \centering{IoT} & Blockchain can provide an infrastructure of distributed devices that replicates the data and validates transactions through secure contracts~\cite{andersen2017wave,zhang2017iot,novo2018blockchain, manzoor2018blockchain,christidis2016blockchains}. \\ \hline
		
		\multirow{2}{*}{\textbf{Logistic}} & \centering{Delivery contract} & Smart contracts can help suppliers obtain anonymous information about customer stock levels, demands and future outlooks in real time, so that it is able to regulate its own production and meet the demands~\cite{hackius2017blockchain,altawy2017lelantos}.\\ \cline{2-3}
		& \centering{Package delivery} & Allows the customer, merchant, and a set of customer-chosen delivery companies to engage in a delivery agreement. In this case, a smart contract acts as a trusted intermediary to enforce fair monetary transactions and enable the communications between contractual parties~\cite{alvarez2017smart,hasan2018blockchain}. \\ \hline
		
		\specialrule{.12em}{0em}{0em}
	\end{tabular}
	\caption[]{Smart contract applications} 
	\label{tab:applications}
\end{table}

\section{Research and Open Challenges}
\label{sec:challenges}
Although smart contracts have tremendous potential in solving real-life problems, most existing platforms and applications are still in their preliminary stage. Common problems smart contracts face range from semantic dependencies to the pseudonymous operation of criminal activities. In this section, we analyze limitations of existing smart contracts and solutions proposed in recent research studies, identify remaining challenges and provide insights on future directions. 
We categorize these challenges into three main classes, namely \emph{technology}, \emph{legalization} and \emph{usability and acceptance}.

\subsection{Technology}
We discuss below the weak links and challenges in the composition and execution of smart contracts from a technical perspective. Note that we here only provide a limited number of examples, a more detailed mapping study on various issues of smart contracts can be found in \cite{alharby2017blockchain}.

\subsubsection{Security}
Security is one of the major concerns of any blockchain system and related procedure. In 2016, a re-entrancy attack in Solidity caused a loss over 40M USD and has led to a heated discussion over security issues of Etheruem smart contracts. In fact, many vulnerabilities are caused by the misunderstanding of the scripting languages \cite{atzei2017survey}.

Following the study conducted by Juels et al. \cite{juels2016ring} in which several forms of criminal Ethereum smart contracts were explored, Luu et al. \cite{luu2016making} further studied security flaws of existing Ethereum smart contracts including how contract execution and code behaviour are affected by the order of mined transactions, correctness of time-stamps and handling of exceptions. Delmolino et al. summarized common mistakes students made while programming smart contracts in the Serpent language \cite{delmolino2016step}. Apart from not realizing the limitation of the blockchain implementation, Delmolino et al. found that students often fail to encode state machines logically and ensure the incentive compatibility of a contract. Wang et al. \cite{wang2019blockchain} categorized semantic vulnerabilities of smart contracts into transaction-ordering dependence, time-stamp dependence, mishandled exceptions, re-entry attacks and call-stack depth.

To enhance security of smart contracts, Luu et al. developed \emph{OYENTE} for to analyzing and detecting security-related document bugs of smart contracts and proposed a set of improvements to the Ethereum protocol. Similarly, Securify~\cite{securify} and Mythril~\cite{mythril} are also intended to ensure security of smart contracts.
Some other groups are also developing alternatives. For instance, the Obsidian coin, developed by Coblenz et al. \cite{coblenz2017obsidian}, comes with a new programming language to enhance the security and usability of smart contracts.
The improvement of existing smart contract languages and development of new ones should be carefully examined. Also, since the types of attacks vary from platform to platform, there is a need to understand the mechanism and vulnerabilities of particular blockchain platforms before using them.

\subsubsection{Privacy}
The pseudonymity of public smart contract do not necessarily guarantee their privacy. In particular, they do not guarantee unlinkability, which is crucial not only for privacy but also for fungibility~\cite{meiklejohn2013fistful}.

One way to protect privacy is to integrate an extra component for data protection, e.g., the Zero-Knowledge Proofs (ZKP) scheme as in ZeroCoin \cite{miers2013zerocoin}. Similar ideas and techniques have also been applied to smart contracts. In Hawk~\cite{kosba2016hawk}, a privacy-preserving compiler was built on top of the ZeroCoin protocol to enable the compilation of smart contracts with a cryptographic protocol while maintaining users' on-chain privacy and contractual security. With a minimally-trusted manager who executes the code, two users can perform actions on smart contracts without revealing the actual information. Another branch of research is around coin mixing. 
For example, CoinShuffle~\cite{ruffing2014coinshuffle} hides the origin of transactions among a group of users by allowing them to shuffle freshly generated output addresses in an oblivious manner.
Similar proposals include ValueShuffle~\cite{ruffing2017mixing} and CoinJoin~\cite{maurer2017anonymous}. However, the adoption of encryption algorithms often brings extra computational overhead for the system, hence future development of privacy preserving techniques shall target light-weight solutions.

\subsubsection{Integrity}
Although the execution of smart contracts is regulated by hard-coded software programs and performed by all network participants, the data fed to smart contracts is still controlled by outside parties and cannot be fully trusted.

Town Crier by Zhang et al. \cite{zhang2016town} serves as a bridge between smart contracts and popular websites to secure the data-delivery. Deployed on the Intel Software Guard Extensions (SGX) hardware that provides a secure enclave for software processing, Town Crier can reliably fetch data from trusted websites to blockchain smart contracts, however, it does not ensure the integrity of data fed towards users. In most cases, users cannot directly access data on a blockchain or smart contract. Instead, they do so via wallet apps developed by other parties, which makes data integrity out of users' control. 

\subsection{Legalization}
Before permissioned smart contracts become ready for a wider adoption in business procedures, many fundamental issues are yet to be solved. Notably, there is still lack of formalized ways of composing smart contracts to suit various design purposes, especially when legal contents are involved. From a legal perspective, there is lack of regulation and policies over smart contracts. It is sometimes hard for blockchains and smart contracts to obtain government approval. By now there is still the issue of enforceability and jurisdiction with this technology. When evaluating opportunities, organizations should carefully evaluate the effect of such lack of government acceptance.

Scripting languages need to be regulated in a way to be more comprehensive and easy-to-use for both technical and non-technical people. In the case of Solidity, Frantz et al. \cite{frantz2016institutions} have proposed a reasonable way of mapping contractual semantics to software declarations that covers the 5 essential components, i.e. "Attributes", "Deontic", "Aim", "Conditions" and "Or else" (or "ADICO"). According to the authors, to successfully convert between institutional constructs and smart contracts, both directions need to be taken into consideration \cite{frantz2016institutions}. 

\subsection{Usability and Acceptance}
\subsubsection{Usability}
Smart contracts as logic-based computer programs have a limited level of interactivity and do not allow people to negotiate and make changes based on the later agreed modifications like in traditional contracts, and they are not flexible with exceptions such as glitches. Also, due to the P2P nature of blockchains, letting ordinary users control their data directly is risky, and the exchange rate can be unpredictable when crypto-currencies are involved.

\subsubsection{Acceptance}
Despite the hype of blockchains and smart contracts in both public and consortium domains, there are still a number of misconceptions about the technology. Firstly, there have been an inflated expectation and many unrealistic use cases. Secondly, even with proper use cases, it can be hard to persuade stakeholders and users to accept the new technology. This could result in extra development costs and a low return on the investment. 

Some of the proposed use cases are in fact more efficient to implement via traditional databases. Hence, those who are interested in developing smart contract applications should keep in mind what can be achieved and what can not with it, as well as the development cost. \\

Further, a summary of applications and challenges associated with them are listed in Table~\ref{tab:challenges}.

\begin{sidewaystable}[!ht]
	\tiny
	\centering
	\begin{tabular}{|p{.2cm}| m{3cm}| m{4cm}| m{.2cm}| m{.2cm}| m{.2cm}| m{.2cm}| m{.2cm}| m{.2cm}| m{.2cm}|m{.2cm}| m{.2cm}| m{.2cm}| m{.2cm}| m{.2cm}| m{.2cm}| m{.2cm}| m{.2cm}| m{.2cm}| m{.2cm}| m{.2cm}| m{.2cm}| m{.2cm}|} \specialrule{.12em}{1em}{0em}
		
		\multicolumn{2}{|c|}{\centering{\textbf{Challenge}}} & \centering{\textbf{Description}} & \rotatebox{90}{\textbf{Banking}} & \rotatebox{90}{\textbf{Know Your Customer (KYC)}} & \rotatebox{90}{\textbf{Insurance}} & \rotatebox{90}{\textbf{Bonds}} & \rotatebox{90}{\textbf{Delay-tolerent micro-payment schemes}} & \rotatebox{90}{\textbf{Charity}} & \rotatebox{90}{\textbf{Electronic Medical Record (EMR)}} & \rotatebox{90}{\textbf{Population health data access}} & \rotatebox{90}{\textbf{Patient matching and identification}} & \rotatebox{90}{\textbf{Personal health tracking}} & \rotatebox{90}{\textbf{Provenance on supply chains}} & \rotatebox{90}{\textbf{Voting}} & \rotatebox{90}{\textbf{Commercial Real Estate (CRE)}} & \rotatebox{90}{\textbf{Resource sharing}} & \rotatebox{90}{\textbf{Home automation}} & \rotatebox{90}{\textbf{Automotives}} & \rotatebox{90}{\textbf{Mobile networks}} & \rotatebox{90}{\textbf{Internet of Things (IoT)}} & \rotatebox{90}{\textbf{Package delivery}} & \rotatebox{90}{\textbf{Smart grids}} \\ \hline 
		
		\multirow{4}{*}{\rotatebox{90}{\textbf{Technology}}} & \centering{Security} & \centering{Introduction of new threat vectors and security vulnerabilities, such as hacks of smart contracts and escape hatches.} & X & X & X & X & X & X & X & X & X & X & X & X & X & X & X & X & X & X & X & X\\ \cline{2-23}
		
		& \centering{Privacy} & \centering{Lack of secrecy in contract execution and user information.} & X & X & & & & X & & & X & & X & & X & X & X & & & & X & \\ \cline{2-23}
		
		& \centering{Integrity} & \centering{When smart contracts fetch data from the web, there's no guarantee on the data integrity.} & & & & & & & & & & & & & & & & & & & & \\ \hline
		
		
		\multirow{2}{*}{\rotatebox{90}{\textbf{Legacy}}} & \centering{Regulations, common industry standards, legal frameworks} & \centering{The legal and regulation frameworks should be defined and accepted for each application domain, currently there are no organizations to standardize the technology.} & X & X & X & X & & & X & X & X & X & X & & & & & & X & X & & \\ \cline{2-23}
		
		& \centering{Lawful recognition and intervention} & \centering{Ability of lawful organizations to recognize smart contract terms and get involve in solving disputes and spams.} & X & & X & X & & & & & & & X & X & X & & & & & & X & \\ \hline
		
		\multirow{4}{*}{\rotatebox{90}{\textbf{Usability}}} & \centering{Flexibility} & \centering{Smart contracts are computer programs that cannot respond well to glitches.} & & & & & & & & & & & & & & & & & & & & \\ \cline{2-23}
		
		& \centering{Risk of user controlling data} & \centering{Lack of knowledge of ordinary users can jeopardize the data.} & & X & & & & & X & & & X & & & & & & & & X & & \\ \cline{2-23}
		
		& \centering{High fluctuation in exchange rate/high liquidity} & \centering{The value of digital assets including currencies is changing fast.} & X & & & X & & X & & & & & X & & & & & & X & X & & X \\ \cline{2-23}
		
		\multirow{3}{*}{\rotatebox{90}{\textbf{Acceptance}}} & \centering{Conceptual misalignment and possible resistance from stakeholders} & \centering{Different opinions on the use of smart contracts of different stakeholders, and the reluctance from some of them to accept a new technology due to the lack of knowledge and potential financial risks.} & X & X & X & X & & X & & X & & X & X & X & & & & & & & & \\ \cline{2-23}
		
		& \centering{Value proposition, extra development costs} &  \centering{Extra budget on deployment with low return of investment.} & & & & X & X & & & & X & & & & X & X & & X & X & & & \\ \cline{2-23}
		
		& \centering{General users} & \centering{Most ordinary users are not aware of the new technology and not willing to participate.} & & & & & X & X & X & X & X & X & & & & X & & & & & X & \\ \hline
		
		
		\specialrule{.12em}{0em}{0em}
	\end{tabular}
	\caption[]{Remaining challenges}
	\label{tab:challenges}
\end{sidewaystable}

\section{Conclusion}
\label{sec:conclusion}
Smart contracts are gaining an increasing popularity in both public and private domains as they enable peer-to-peer operation on public blockchains and have the potential to improve efficiency and transparency in business collaborations. However, the current form of smart contracts are still limited in their ability to full fill all expectations. We believe the future development should mainly focus on improving semantics of smart contracts, their integration with existing procedures, as well as their usability, acceptance and legality. If smart contracts can be made to work with enhanced security, legality and flexibility, we can foresee a wider adoption of smart contracts.

\bibliographystyle{abbrv} 
\bibliography{bibliography}
	
\end{document}